\documentclass[final,3p,times,12pt]{elsarticle}
\usepackage{color}
\usepackage{graphicx}
\usepackage{subfigure}
\usepackage{slashed}
\usepackage{amsmath}

\newcommand{\code}[1]{{\small{\texttt{#1}}}}
\newcommand{\delphi}{\ensuremath{\Delta \hat{\phi}}}

\usepackage[colorlinks=true
,urlcolor=blue
,anchorcolor=blue
,citecolor=blue
,filecolor=blue
,linkcolor=black
,menucolor=blue
,pagecolor=blue
,linktocpage=true
]{hyperref}

\usepackage{amssymb}

\journal{Physics Letters B}

\begin{document}
\begin{flushright}
 OHSTPY-HEP-T-14-001
\end{flushright}
\begin{frontmatter}
\title{Connecting Simplified Models: Constraining Supersymmetry on Triangles}
\author[label1]{Archana Anandakrishnan}
\ead{archana@physics.osu.edu}
\author[label1]{Christopher S. Hill}
\ead{chill@physics.osu.edu}
\address[label1]{Department of Physics, The Ohio State University,
191 W.~Woodruff Ave, Columbus, OH 43210, USA}

\begin{abstract}
We investigate an approach for the presentation of experimental constraints on supersymmetric scenarios. It is a triangle based visualization that extends the status quo wherein LHC results are reported in terms of simplified models under the assumption of 100\% branching ratios.  We show that the (re)interpretation of LHC data on triangles allows the extraction of accurate exclusion limits for a multitude of more realistic models with arbitrary branching ratios. We demonstrate the utility of this triangle visualization approach using the example of gluino production and decay in several common supersymmetric scenarios.

\end{abstract}


\end{frontmatter}

\section{Introduction}
\label{sec:intro}
The extent to which supersymmetric scenarios are excluded by data from the LHC experiments is obscured by the breadth of realizations with which such scenarios might manifest themselves.  In order to make this problem more tractable while avoiding the prejudices of specific UV completions (e.g. CMSSM), the ATLAS and CMS experiments have adopted the strategy of distilling theoretical scenarios into ``Simplified Models"~\cite{Alwall:2008ag, oai:arXiv.org:1105.2838, Chatrchyan:2013sza} that reduce the parameters of the theory to those that directly affect the experimental observability of the supersymmetric signal.  While this has been a significant improvement in the way LHC experimental constraints on supersymmetry are presented, this approach also has a number of shortcomings.  In this Letter, we present an extension of the simplified model approach that addresses one such limitation, namely the commonly made, though often unrealistic assumption that any new particles produced will have 100\% Branching Ratio (BR) into the experimental final states over which the search is conducted. 

In this work we will focus on the gluino as an example of a SUSY particle where the existence of a number of possible decay modes complicates the interpretation of the experimental results produced by the LHC collaborations. Even in scenarios in which the gluino decays only to the lightest neutralino plus a quark-antiquark pair, there are distinct possibilities for the decay that are optimized with different search strategies. In principle, the possibility of two different decay modes for a pair-produced supersymmetric particle could significantly weaken the exclusion limits obtained by assuming 100\% BR into the final states considered in the experimental search.  The triangle approach we adopt here for presentation of the experimental limits allows one to visualize this effect.

\section{Points on the Triangle}
\begin{figure*}[ht!]
\begin{center}
 \includegraphics[width=9.0cm]{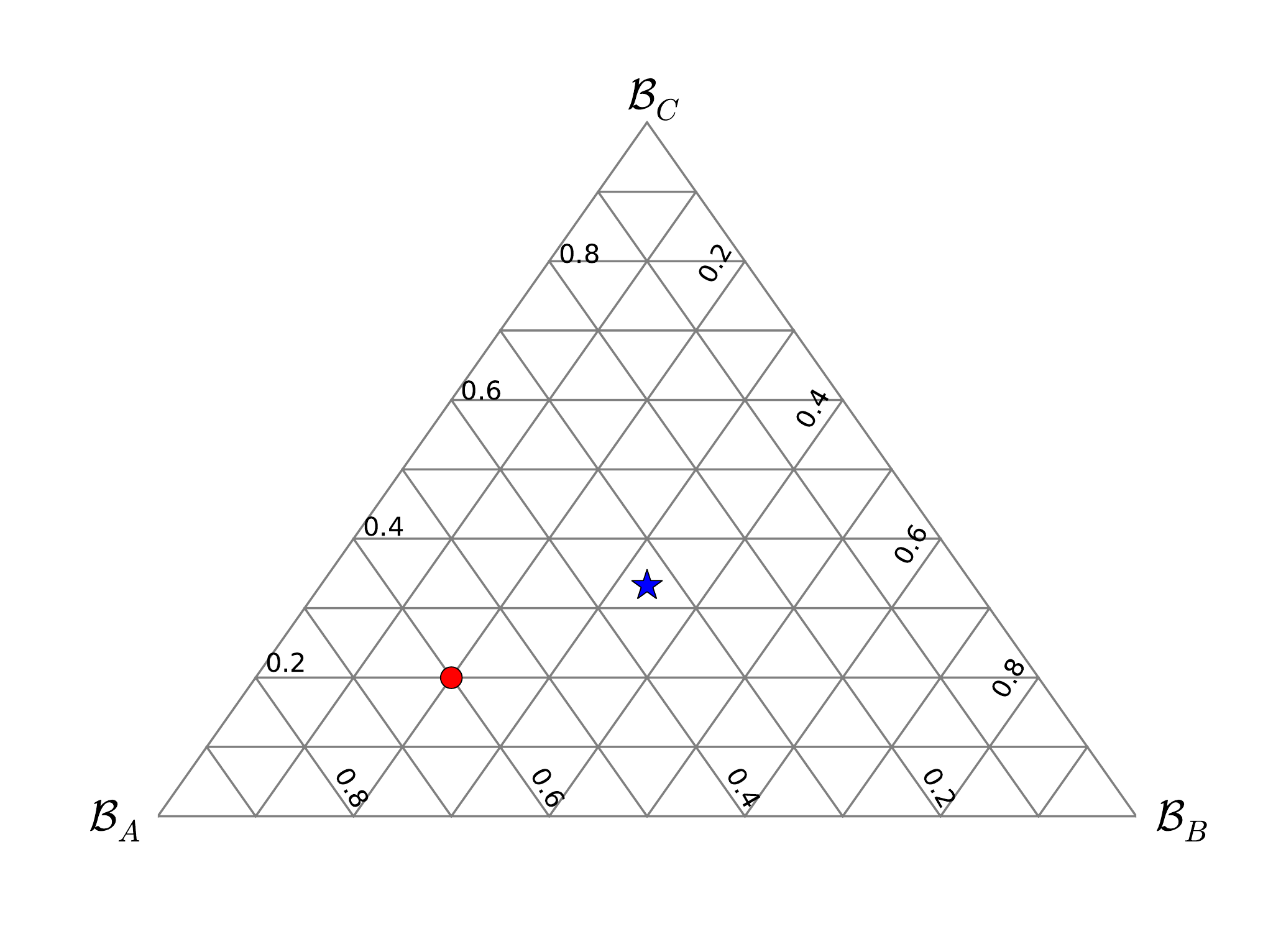}
\caption{\label{skeleton} Every point in this skeleton grid has a unique value for branching ratios ${\mathcal B}_A$, ${\mathcal B}_B$, and ${\mathcal B}_C$. Each particular branching ratio decreases from 1 at a vertex to 0 at the side opposite to that vertex. The grid lines are  drawn to show the variation of the branching ratios in each direction inside the triangle. The point marked by a {\color{blue} $\bigstar$} denotes the centroid of the triangle and is composed of equal branching ratios, $({\mathcal B}_A, {\mathcal B}_B, {\mathcal B}_C) = (33\%,33\%,33\%)$. The point marked by {\color{red} $\bullet$} is composed of branching ratios, $({\mathcal B}_A, {\mathcal B}_B, {\mathcal B}_C) = (60\%, 20\%, 20\%)$. }
\end{center}
\end{figure*}

By definition, the sum of a particle's branching ratios add up to one. A particle with 100\% BR into a single set of final states represents a Simplified Model Scenario (SMS) which is a single point on the parameter space of all possible models of the particle. Models with branching ratios of a particle into two independent final states all lie on a straight line given the constraint on the total branching ratio. Similarly, all models of the particle with three independent decay modes are confined to a triangle since there are only two free parameters. Note that models with greater than 3 decay final states cannot be visualized in the same manner on a 2-dimensional plot. Nevertheless, all models with upto three decay final states can be presented by adopting the triangle visualization method. Let us denote the three decay branching fractions of a supersymmetric particle as ${\mathcal B}_A$, ${\mathcal B}_B$, and ${\mathcal B}_C$. The space spanned by scanning over values of (${\mathcal B}_A$, ${\mathcal B}_B$, ${\mathcal B}_C$) is a triangular plane. Each point on the 2-dimensional space can be written in terms of the branching ratios as:
\begin{equation}
 (x,y) = \left( \frac{1}{2} \frac{2{\mathcal B}_B + {\mathcal B}_C}{{\mathcal B}_A + {\mathcal B}_B + {\mathcal B}_C}, 
\frac{\sqrt{3}}{2} \frac{{\mathcal B}_C}{{\mathcal B}_A+{\mathcal B}_B+{\mathcal B}_C} \right)
\end{equation}
With the constraint that ${\mathcal B}_A + {\mathcal B}_B + {\mathcal B}_C = 1$, the three vertices of the equilateral triangle (with the vertices located at $(0,0)$, $(1,0)$ and at $(\frac{1}{2}, \frac{\sqrt{3}}{2})$) correspond to simplified models with ${\mathcal B}_A = 1$, ${\mathcal B}_B = 1$, and ${\mathcal B}_C =1$ respectively. The skeleton grid representing each point on the 2-dimensional plane is shown in Fig.~\ref{skeleton}. The grid lines serve as a guide to read off the composition of branching ratios at each point within the triangle. For example, the point marked by a {\color{blue} $\bigstar$} denotes the centroid of the triangle and is composed of equal branching ratios, $({\mathcal B}_A, {\mathcal B}_B, {\mathcal B}_C) = (33\%,33\%,33\%)$. Similarly, the point marked by {\color{red} $\bullet$} is composed of branching ratios, $({\mathcal B}_A, {\mathcal B}_B, {\mathcal B}_C) = (60\%, 20\%, 20\%)$. 

In this representation, the vertices are the Simplified Model scenarios (SMS) for which experimental constraints have been published in the literature. The edges connecting any two vertices of the triangle are models with various combinations of the two branching ratios located at the respective vertices. The rest of the triangle, however, contains more realistic scenarios with multiple decay modes of the parent particle that have not been explicitly confronted with LHC data (though in principle they could be).  In fact, the advantages of such triangular visualization have already been demonstrated in two non-supersymmetric BSM analyses, namely the search for vector-like top partners, ``T quarks"~\cite{Chatrchyan:2013uxa, ATLAS-CONF-2013-018}. Since a T quark can decay into three final states: $bW$, $tZ$ or $tH$, by presenting the experimental constraints on triangles, CMS and ATLAS were able to conduct the search without making any assumptions on the branching fractions and place the most stringent bounds on the entire parameter space.  In this work, we apply the triangle visualization approach to searches for supersymmetry where it will be useful, given the fact that most supersymmetric particles decay into multiple final states as stated in the Introduction.

\section{Example: Gluino Decays}
In Ref.~\cite{Chatrchyan:2013wxa}, the CMS collaboration searched for evidence of supersymmetry in events with large missing energy, jets, b-jets and no leptons. In addition, events were required to have $\Delta \hat{\phi}_{\text{min}} > 4.0$, where $ \Delta \hat{\phi}_{\text{min}} = \text{min} \left( \Delta \phi_i/\sigma_{\Delta \phi_i} \right) $ and $\Delta \phi$ is the angle between a jet and the negative of the $\slashed{E}_T$ vector, and $\sigma_{\Delta \phi_i}$ is the estimated resolution of $\Delta \phi$. By requiring $\Delta \hat{\phi}_{\text{min}} > 4.0$, most of the QCD background was eliminated. The observed number of events in several signal regions was consistent with the expected SM backgrounds and 95\% upper limits on the presence of new physics were extracted. The upper limits were interpreted as bounds on gluino production in two different SMS: (i) 100 \% branching ratio of $\tilde{g} \rightarrow b \bar{b} \tilde{\chi}^0_1$ (T1bbbb) and (ii) 100 \% branching ratio of  $\tilde{g} \rightarrow t \bar{t} \tilde{\chi}^0_1$ (T1tttt). By requiring b-jets and vetoing events with leptons, the search was most sensitive to the T1bbbb SMS. At 95\% C.L., gluinos lighter than 1170 GeV were excluded in the T1bbbb SMS and gluinos lighter than 1050 GeV in the T1tttt SMS. In this section, we will show that the results from the ``\delphi" analysis in Ref.~\cite{Chatrchyan:2013wxa} can be reinterpreted to obtain stringent constraints on a wide range of more realistic models with the triangle visualization approach. 

We pick three branching ratios and consider benchmarks points along the grid in Fig.~\ref{skeleton}. For each point, the procedure is similar. We generate 10,000 gluino pair-production events for 8 TeV LHC using \code{PYTHIA 8.175}~\cite{Sjostrand:2007gs}. \code{PYTHIA} decays the gluino pair according to the branching ratios at the point on the triangle and hadronizes the decay products.  Next, a detector simulation is employed to estimate the Acceptance $\times$ Efficiency ($A \times \varepsilon$) for these signal events to pass the selection criteria of Ref.~\cite{Chatrchyan:2013wxa}. While the most accurate estimation of $A \times \varepsilon$ can only be achieved by a full \code{GEANT}~\cite{Agostinelli:2002hh} simulation of the detector (which can only be performed by the experimental collaborations), a decent parametric simulation of the detector is sufficient for our purposes and for this we use \code{Delphes 3.0.9}~\cite{deFavereau:2013fsa}.  We use the default ``CMS" detector card provided by \code{Delphes} adapted to account for the electron and muon isolation criteria applied in the CMS analysis and modified to match the track and jet reconstruction parameters quoted in Ref.~\cite{Chatrchyan:2013wxa}. In addition, we modify the b-tagging efficiency in the \code{Delphes} CMS detector card using the efficiency information for the combined-secondary-vertex algorithm reported in Ref.~\cite{CMS-PAS-BTV-11-004}. 

\begin{figure*}[ht!]
\centering
\includegraphics[width=0.70\textwidth]{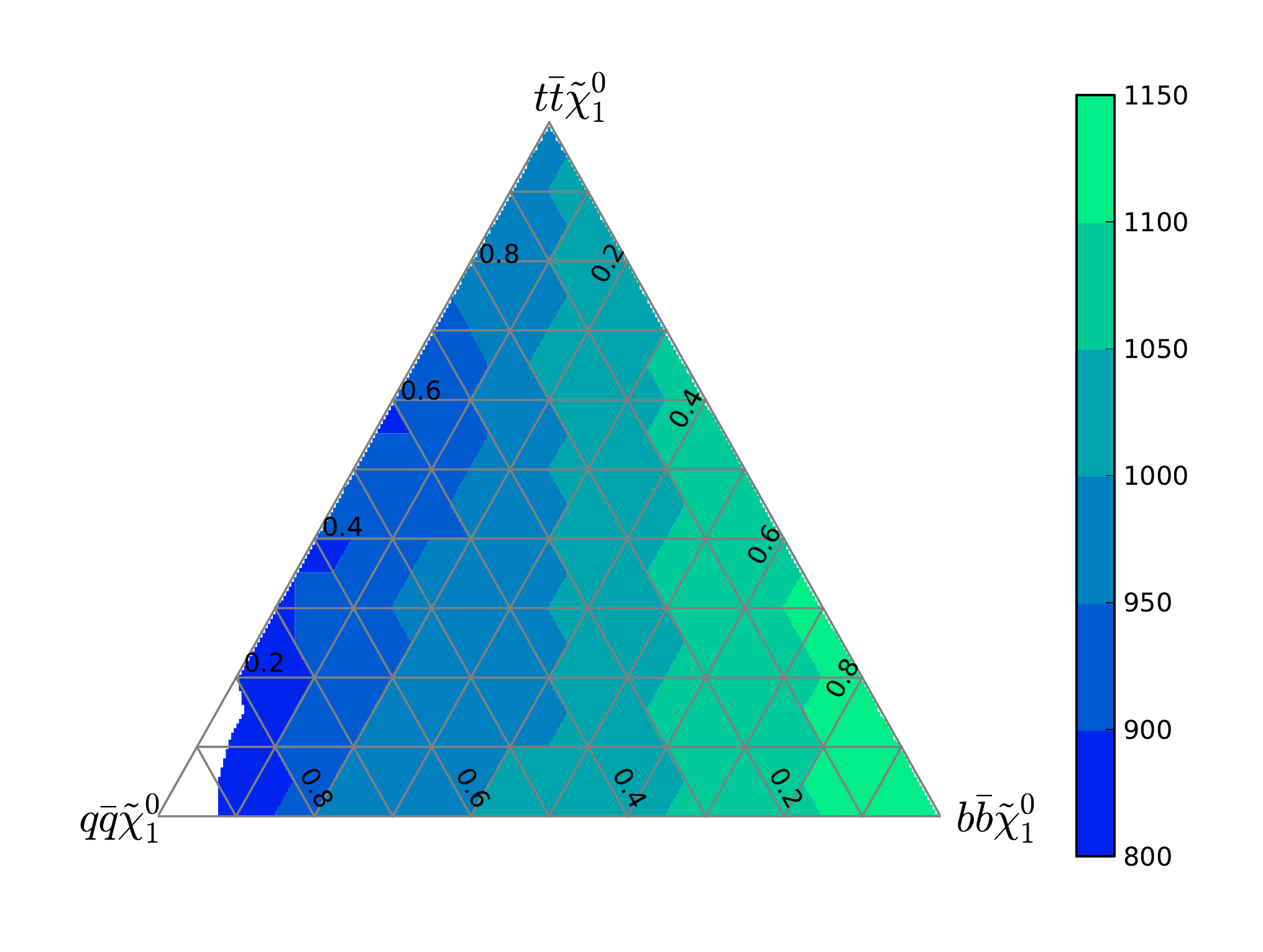}
\vspace{-1cm}
\caption{\label{fig:FG-1}The interpretation of CMS analysis Ref.~\cite{Chatrchyan:2013wxa} for benchmark points with various combinations of decay branching fractions of the gluino to $b \bar{b} \tilde{\chi}^0$, $t \bar{t} \tilde{\chi}^0$ and $q \bar{q} \tilde{\chi}^0$. Each point in the above triangle has a unique combination of the three branching fractions and the vertices represent the simplified models with 100\% branching fractions into one of the three final states. We do not show gluino mass limits in the region near the $q \bar{q} \tilde{\chi}^0$ vertex since the limits on models with large decays to $q \bar{q} \tilde{\chi}^0$ are very weak and would be off-scale.}
\end{figure*}

Finally, our dedicated C++ code reads in the \code{ROOT} file output from \code{Delphes} and implements the event selection from Ref.~\cite{Chatrchyan:2013wxa}. The events that pass the selection criteria in each signal region are scaled by the appropriate NLO cross-section~\cite{Kramer:2012bx} and normalized to an integrated luminosity of 19.6 fb$^{-1}$ to be consistent with the published analysis. The expected number of signal events, $N_{\rm sig} = \sigma_{\rm NLO} \times \int\mathcal{L} dt \times (A \times \varepsilon)$, obtained in this manner are compared to the observed number of events quoted by the experimental collaboration in their published analysis~\cite{Chatrchyan:2013wxa}.  A specific model is considered excluded if $N_{\rm sig} > N_{\rm UL}$ where $N_{\rm UL}$ is the 95\% Bayesian upper limit (assuming a flat prior) on events produced by the BSM process, computed given the estimated SM backgrounds and the observed number of events, reported in Ref.~\cite{Chatrchyan:2013wxa}.  Using this procedure, we fill the skeleton grid shown in Fig.~\ref{skeleton} with color maps of the gluino exclusion limit at 95\% C.L. from the best signal region of the \delphi\ analysis and the results are presented in Figs.~\ref{fig:FG-1},~\ref{fig:FG-2} and~\ref{fig:FG-3}. There is a kinematic lower bound on the gluino mass for each model.  In addition, the experimental efficiencies degrade rapidly for small mass differences between the gluino and LSP due to the relatively low momenta of the decays products.   Consequently, the bounds do not extend all the way to zero-mass of the gluino and we consider the range of gluino mass specified by the analysis, which in this case is 400 GeV and above. 

The first result we present is for the simplest scenario where only one neutralino is lighter than the gluino. In this case, the gluino predominantly decays into $q \bar{q} \tilde{\chi}^0_1$, $b \bar{b} \tilde{\chi}^0_1$, $t \bar{t} \tilde{\chi}^0_1$. The triangle with the branching fractions for these decay modes set to 100\% at the vertices, with the neutralino mass set to 100 GeV and all other supersymmetric particles decoupled is shown in Fig.~\ref{fig:FG-1}.  Note, we can check our \delphi\ reinterpretation analysis against the CMS analysis at the vertices that correspond to the T1bbbb and T1tttt simplified model scenarios. At these two vertices, we find that gluinos lighter than 1200 and 1080 GeV are excluded respectively.  These results agree with those published in Ref.~\cite{Chatrchyan:2013wxa} within the uncertainties of the detector simulation and we therefore consider our procedure, and in  particular our use of \code{Delphes} to estimate $A \times \varepsilon$, to be validated. The triangle visualization approach, however, also contains significant additional information about the effect of the mixed final states on the model constraints that can be extracted from the same data. For instance, from Fig.~\ref{fig:FG-1} we can immediately see that with about 65\% branching ratio into $b \bar{b} \tilde{\chi}^0_1$ and 35\% branching ratio into $t \bar{t} \tilde{\chi}^0_1$, the limits are as strong as the T1bbbb SMS. An additional benefit that the triangle visualization approach provides can be seen by allowing the third vertex of the triangle to determine the constraints in models with the least efficiency for signal events to pass the kinematic selection criteria. For example, events from gluino decays into $q \bar{q} \tilde{\chi}^0_1$, where $q = u,d,c,s$ are less likely to pass the $b$-tagged jet selection criterion in the \delphi\ analysis since the final states do not contain any $b$ quarks. Consequently, the simplified model T1qqqq has weaker bounds on it coming from this analysis, but it is also evident from Fig.~\ref{fig:FG-1} that with only 20\% decay branching ratio into $b \bar{b} \tilde{\chi}^0$, the constraints on the gluino mass very rapidly increase to 1000 GeV. By studying final states that have very different efficiencies of populating the signal regions on a single triangle, one can constrain a wide range of models with a given experimental analysis.

Let us look at some additional specific examples to further illustrate the utility of the triangle visualization approach.  Consider model A with equal gluino branching ratios into $q \bar{q} \tilde{\chi}^0_1$, $b \bar{b} \tilde{\chi}^0_1$, $t \bar{t} \tilde{\chi}^0_1$ ({\color{blue} $\bigstar$} in Fig.~\ref{skeleton}).  By locating this model at the centroid of the triangle in Fig.~\ref{fig:FG-1} one can obtain an accurate lower bound of 1050 GeV on the gluino mass. Now consider model B that has 20\% branching ratio into $b \bar{b} \tilde{\chi}^0_1$, 20\% into $t \bar{t} \tilde{\chi}^0_1$, and the remaining 60\% into states that are not shown on this triangle. We can still obtain a lower bound on the gluino mass of at least 1000 GeV by locating the model {\color{red} $\bullet$} on the triangle in Fig.~\ref{fig:FG-1}. Since this bound is obtained by assuming that the remaining 60\% of the gluino decay products would have significantly less efficiency of passing the selection criteria of the dedicated experimental analysis, the actual bound would be between 1000 - 1200 GeV.  And lastly, consider model C with 30\% branching ratio to $b \bar{b} \tilde{\chi}^0_1$ and 70\% branching ratio into final states that are also not shown on the triangle.  In this case, we can still infer that the given experimental search excludes gluino masses between 1000 - 1200 GeV from the grid line corresponding to 30\% decays into $b \bar{b} \tilde{\chi}^0_1$. The bounds obtained for models B and C assume that $A \times \varepsilon$ for the modes not pictured in the triangle are equal or larger than those for T1qqqq.  This assumption is valid since, by construction, the third vertex was chosen to be a decay mode with very small efficiency of passing the cuts. For the model T1qqqq, we find that $A \times \varepsilon < 10^{-4}$. Therefore, this construction enables us to obtain bounds for models even when all the BRs are not pictured on the triangle.  

Currently the ATLAS and CMS collaborations obtain exclusion limits in the context of Simplified Model Scenarios. Exclusion curves are projected on a 2-dimensional $m_{\rm LSP} - m_{\tilde{g}}$ parameter space. The exclusion limits depend on the mass difference between the gluino (or any other supersymmetric particle in study) and the LSP and typically the limits are strongest  when the mass difference is large. Nevertheless, the SMS is a single point in the model parameter space and by extending the model space to include up to three decay final states we sacrifice information on how the limits change with the mass difference. The combination of the SMS presentation and triangle plots provide all necessary information such as the kinematic reach of the analysis, consistency checks between the expected and observed limits and in addition, information in the model space with varied branching ratios.

\section{Other Models on Triangles}
In the previous section, we demonstrated the potential and the breadth of the triangle visualization approach by considering the simple example of gluino decays into the lightest neutralino and a variety of SM final states. Many models have very specific predictions for the decay branching ratios of the gluino (and other supersymmetric particles). And typically, more than one of the electroweakinos are lighter than the gluino and cascade decays of the gluino become prominent~\cite{Alves:2011sq}. We show that it is possible to use the triangle visualization approach to present constraints for such models by examining two illustrative examples. 

\begin{figure*}[ht!]
\centering
\subfigure{
\includegraphics[width=0.48\textwidth]{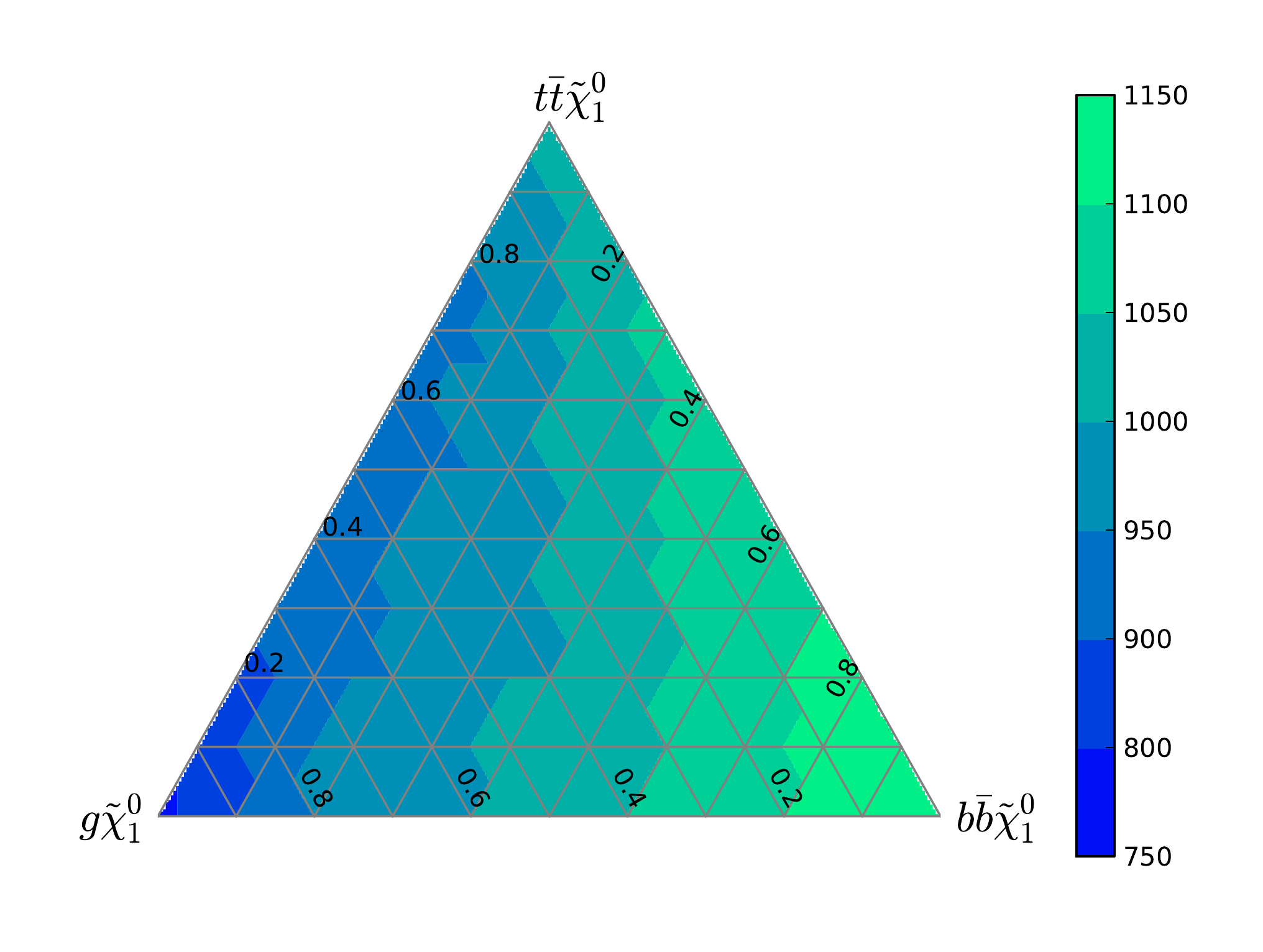}
\label{fig:FG-2}
}
\subfigure{
\includegraphics[width=0.48\textwidth]{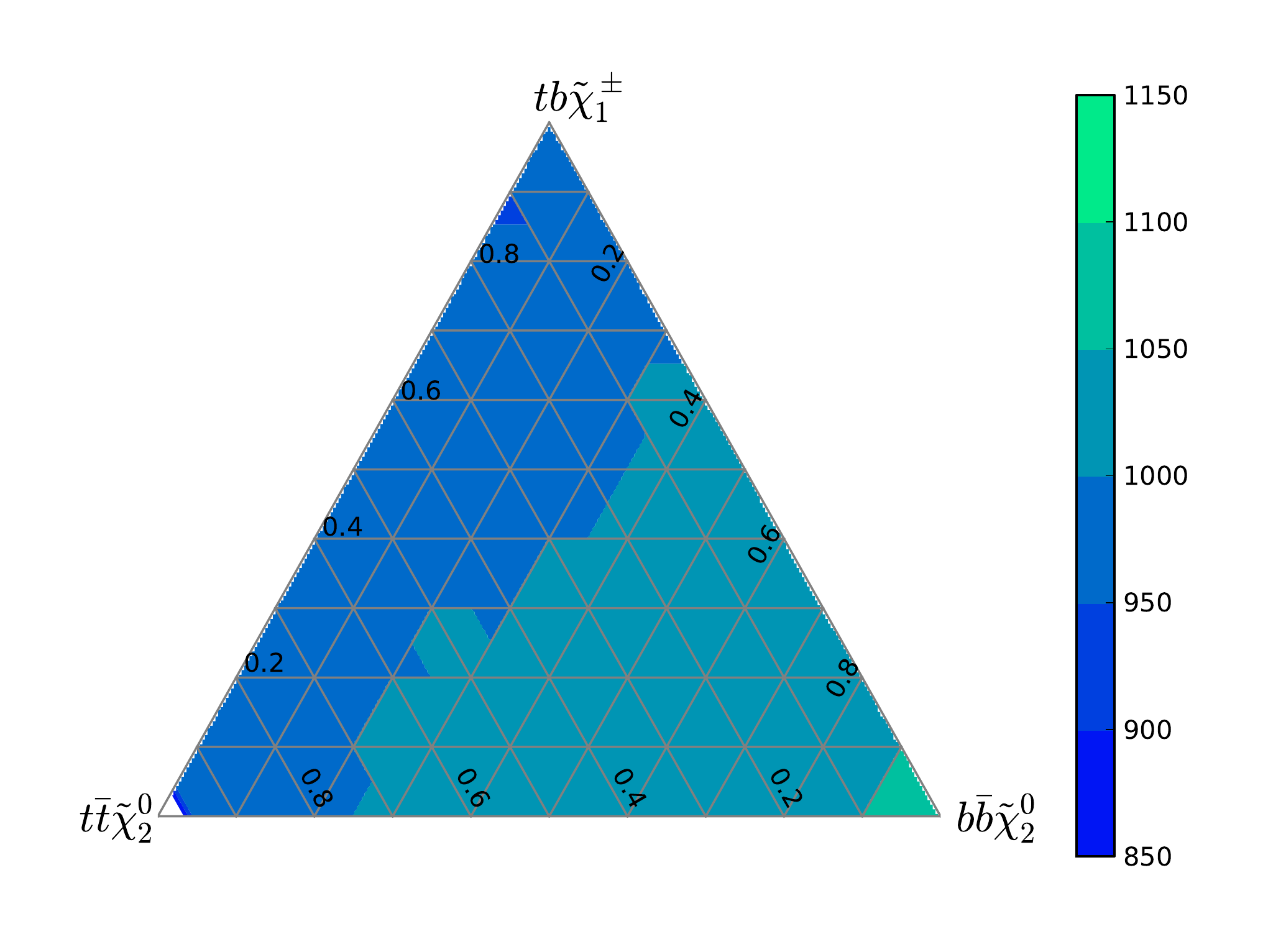}
\label{fig:FG-3}
}
\caption{The reinterpretation of the CMS analysis Ref. \cite{Chatrchyan:2013wxa} for two model scenarios. The figure on the left (a) demonstrates the results of the \delphi\ analysis reinterpreted for models with dominant decays into $g \tilde{\chi}^0_1$, $b \bar{b} \tilde{\chi}^0_1$, $t \bar{t} \tilde{\chi}^0_1$. The figure on the right (b) is the reinterpretation of the same analysis with $t b \tilde{\chi}^\pm_1$, $b \bar{b} \tilde{\chi}^0_2$, $t \bar{t} \tilde{\chi}^0_2$ as the vertices of the triangle. }
\end{figure*}

First, consider supersymmetric scenarios with heavy scalars superpartners and lighter gauginos ({\it e.g.} split supersymmetry). In these models, loop decays of the gluino into $g \tilde{\chi}^0_i$ are sizeable due to the large mass of the scalars that mediate the three body decays. In addition, the third family of scalars: the stop, sbottom and stau may be lighter than the scalars of the first and second family, due to renormalization group running. Therefore, decays to $b \bar{b} \tilde{\chi}^0_i$, $t \bar{t} \tilde{\chi}^0_i$ are also present in these scenarios. The results of the \delphi\ analysis reinterpreted for models with heavy scalars and dominant decays into $g \tilde{\chi}^0_1$, $b \bar{b} \tilde{\chi}^0_1$, $t \bar{t} \tilde{\chi}^0_1$ is shown in Fig.~\ref{fig:FG-2}. Notice that the only difference from the triangle shown in Fig.~\ref{fig:FG-1} is the $g \tilde{\chi}^0_1$ decay mode. Here there is sufficient gluon radiation followed by $g\rightarrow b\bar{b}$ such that events pass the $b$-tagged jets criterion of the \delphi\ analysis with reasonable efficiency; hence the constraints are stronger in models with 100\% decays into $g \tilde{\chi}^0_i$, when compared to models with 100\% decays into $q \bar{q} \tilde{\chi}^0_1$.

Similarly, in models with light charginos, the $t b \tilde{\chi}^\pm_1$ decay mode will usually dominate~\cite{Baer:1986au}. In addition, decays to $b \bar{b} \tilde{\chi}^0_2$, $t \bar{t} \tilde{\chi}^0_2$ (or to heavier neutralinos) will also be present if there are other states lighter than the gluino. Although it is impossible to visualize all the different decay possibilities on a single 2-dimensional plot, notice that many of the decays produce very similar final states at the current resolution level of the searches. The decays $t b \tilde{\chi}^\pm_1$ and $t \bar{t} \tilde{\chi}^0_1$, for instance, both produce 2 $W$ bosons, 4 $b$-quarks, and missing energy when $m_{\tilde{\chi}_1^\pm}-m_{\tilde{\chi}_1^0} > m_W$ (Here we assume $m_{\tilde{\chi}_1^\pm}, m_{\tilde{\chi}_2^0} \simeq 2 m_{\tilde{\chi}_1^0}$). Consequently, we obtain the same exclusion limits on models with either of these two decay branching ratios. As another example consider the decays of $t b \tilde{\chi}^\pm_1$, $b \bar{b} \tilde{\chi}^0_2$, $t \bar{t} \tilde{\chi}^0_2$ that produce multiple $b$-jets and missing energy. As a result, from Fig.~\ref{fig:FG-3} we can see that even though they are weaker than the T1bbbb or T1tttt SMS (since a fraction of the $\tilde{\chi}^0_2$, $\tilde{\chi}^\pm_1$ decay into leptons and events with leptons are vetoed by the \delphi\ analysis), the exclusion limits are again similar for all three decays. Overall, the \delphi\ analysis (and most other hadronic searches) has the same sensitivity to models with any combination of these final states and we expect the exclusion limits to be very similar.

Other models can also be constrained using this approach, by projecting three (or more, with additional constraints) branching ratios on a triangle. The simplest triangle to adopt would be the one presented in Fig.~\ref{fig:FG-1}. It is model independent and yet gives the widest range of exclusion limits. The reinterpretation of an experimental analysis for particular models is useful for theorists to guide model building, but for an informative characterization of exclusion limits on supersymmetric particles, the use of the simplest triangle will usually be sufficient. Another utility of the triangle for a specific model with spontaneous R-parity violation in SUSY, was demonstrated by the authors of Ref.~\cite{Marshall:2014cwa, Marshall:2014kea}.  

\section{Conclusion}
We advocate an approach to visualize the results of experimental SUSY searches by presenting these as mass exclusions on a triangle spanning the branching ratio parameter space of a given supersymmetric scenario. With this approach, at the vertices of the triangle experimentalists can still report exclusions on the SMS that are presently being constrained in CMS and ATLAS publications, but one can also present limits on models with mixed final states that are usually omitted in these papers.  This is important since in such models the constraints on supersymmetric particles can be very different from the SMS results depending on the efficiency of the mixed states to populate the signal regions. We also think the triangular visualization will be useful to theorists.  It allows them to use existing published SMS results and interpret them as bounds on supersymmetric particles in specific models that make specific branching ratio predictions.  Examples of this in the context of constraints on the gluino in SUSY scenarios were given.  
We hope we have demonstrated that the triangular visualization approach would significantly improve the utility of the experimental results being reported by the CMS and ATLAS collaborations. Presentation of results in this form, in addition to the usual exclusion limits on the 2-dimensional ($M_{LSP}\ {\rm vs}\ M_{\tilde{g}}$) plot will maximize the information presented and makes it relatively easy to reinterpret the results for particular models with different decay configurations.  We encourage the high energy physics community to consider adopting this presentation strategy as they prepare for Run 2 of the LHC~\footnote{We are pleased to note that since the submission of this draft, the ATLAS collaboration has indeed adopted a similar representation in~\cite{Aad:2014mha}.  However, in this work they have set limits on BR in the triangular plane rather than mass limits as we advocate here (and suggest would be more useful)},~\footnote{A \code{Python} script that can be adapted to visualize data on triangular plots can be obtained from \href{https://github.com/renuk16/Triangles}{https://github.com/renuk16/Triangles}}.

\section*{Acknowledgements}
AA would like to thank Michael Peskin for the suggestion to explore this method. We would also like to thank Stuart Raby and Charles Bryant for useful discussions. We thank the Ohio Supercomputer Center for their computing resources. AA is funded through the Ohio State University Presidential Fellowship.

\section*{References}
\bibliographystyle{elsarticle-num} 
\bibliography{TrianglesDraft}

\end{document}